\documentclass[11pt,a4paper]{article}
\usepackage{jheppub}
\usepackage{stackrel}
\usepackage{epstopdf}

\title{Worldline approach to the Grosse-Wulkenhaar model}
\author{Sebasti\'an Franchino Vi\~{n}as, Pablo Pisani}
\affiliation{Instituto de F\'isica La Plata, CONICET -- UNLP\\
C.C. 67, 1900 La Plata, Argentina}

\emailAdd{safranchino@fisica.unlp.edu.ar; pisani@fisica.unlp.edu.ar}

\date{\today}

\abstract{We apply the worldline formalism to the Grosse-Wulkenhaar model and obtain an expression for the one-loop effective action which provides an efficient way for computing Schwinger functions in this theory. Using this expression we obtain the quantum corrections to the effective background and the $\beta$-functions, which are known to vanish at the self-dual point. The case of degenerate noncommutativity is also considered. Our main result can be straightforwardly applied to any polynomial self-interaction of the scalar field and we consider that the worldline approach could be useful for studying effective actions of noncommutative gauge fields as well as in other non-local models or in higher-derivative field theories.}

\begin{document}

\maketitle

\section{Introduction}

The construction of Feynman diagrams is the basic tool for the perturbative determination of physical quantities in quantum field theories. Even so, a serious limitation of this method is that the number of relevant diagrams increases rapidly with the perturbative order of the quantum corrections or with the number of particles involved in the scattering processes. Nevertheless, in some theories there occur large cancellations between different diagrams that simplify the final results. The worldline formalism (WF) is an efficient method for computing effective actions, scattering amplitudes and anomalies in quantum field theory which explains some of these cancellations and shows many conceptual and practical advantages in relation with the usual diagrammatic procedure \cite{Schubert:2001he,Bastianelli:2006rx}.

In the context of standard field theories, the WF can be used to study the heat-trace and, consequently, the functional determinant of the differential operator whose spectrum describes the quantum fluctuations. However, in noncommutative (NC) field theories \cite{Douglas:2001ba,Szabo:2001kg} nonlocality implies that quantum fluctuations are no longer described by a differential operator. Still, the WF has been recently applied to determine the Seeley-de Witt coefficients in a simple NC model \cite{Bonezzi:2012vr}. These coefficients exhibit some of the renormalization peculiarities of NC theories, which are a consequence of the nonlocality of the interactions.

In fact, it is well-known that the one-loop propagator for a massive scalar field with a quartic self-interaction on the Moyal plane contains contributions --from high-energy virtual excitations-- which diverge for vanishing incoming momenta \cite{Minwalla:1999px}. As a consequence, one-loop corrections to the propagator lead to nonrenormalizable singularities when inserted in higher-loop graphs. In the heat-kernel approach this phenomenon --known as UV/IR mixing and due to the contributions of non-planar Feynman diagrams-- is related to the existence of ``Moyal nonlocal'' Seeley-de Witt coefficients \cite{Vassilevich:2003yz}.

Some years after the discovery of the UV/IR mixing problem, H.\ Grosse and R.\ Wulkenhaar showed that renormalization can be restored by a proper modification of the free propagator of the scalar field \cite{Grosse:2003nw,Grosse:2004yu}. This modification is implemented in the Grosse-Wulkenhaar (GW) model by introducing a harmonic oscillator background, thus breaking translation invariance but restoring Langmann-Szabo duality \cite{Langmann:2002cc}. The resulting Lagrangian has been interpreted as arising from the interaction with the curvature of a properly defined noncommutative space \cite{Buric:2009ss}. The GW model has several remarkable renormalization properties: it is renormalizable to all orders of perturbation theory \cite{Grosse:2003nw,Grosse:2004yu,Rivasseau:2005bh} and it has a vanishing $\beta$-function (at a self-dual fixed point) so it has no Landau ghost \cite{Disertori:2006nq,Rivasseau:2007qda}. Moreover, it is equivalent --at the self-dual point and in the extreme noncommutativity/infinite volume limit-- to an exactly solvable and nontrivial matrix model \cite{Grosse:2012uv}. As a consequence, it stands as a promising framework for a full construction of an axiomatic field theory in four dimensions\footnote{Conditions under which this model could satisfy Osterwalder-Schrader axioms are analyzed in detail in \cite{Grosse:2014lxa}.} \cite{Grosse:2014nza}. Let us also mention that the flow of the GW model turned out to be an interesting scenario for the functional renormalization group approach in which the self-dual model can be studied without using any truncation \cite{Sfondrini:2010zm}.

In the present article we study the Grosse-Wulkenhaar model in the worldline approach. Our purpose is to show that the WF can be applied to this NC model and our main result is a master formula for the one-loop effective action from which the $n$-point Green function can be recovered. From this formula the $\beta$-functions and the effective background are readily obtained. Although the application of the WF to this nonlocal theory encounters some issues which also appear in its application to field theories in curved spaces, its implementation in the GW model does not face the same difficulties and is technically more direct. Our approach can be applied to a scalar field with $\lambda\phi_\star^N$-type self-interaction in $d$-dimensional NC spacetime but we expect that it could also be useful for computing physical quantities in other types of nonlocal theories as well as in field theories with higher derivatives.

The article is organized as follows. In section \ref{GWmodel} we give our notation for the algebra of functions on $d$-dimensional Euclidean Moyal spacetime, we present the GW model for a real scalar field with a quartic self-interaction and we write down the one-loop effective action in terms of the heat-trace of a nonlocal operator. In section \ref{ht} we compute this heat-trace by using a representation in terms of a path integral in the phase space of a point particle.

Afterwards, in section \ref{eaqw}, we use our result for the heat-trace in order to present a master formula --eq.\ \eqref{eff-action-pt2}-- for the effective action of the GW model. Then, as an application, we study the two- and four-point functions. In section \ref{prop}, from planar and non-planar contributions to the propagator, we obtain the field renormalization constant and the frequency $\beta$-function and we write down the finite one-loop contributions to the effective background. Next, in section \ref{fpf}, we study the four-point function in order to determine the coupling constant $\beta$-function. The case of degenerate noncommutativity is considered in section \ref{degen} in some detail. Finally, in section \ref{conclu} we draw our conclusions and in appendix \ref{app1} we show how the WF can also be used to compute local heat-kernels.

\section{Moyal space and Grosse-Wulkenhaar model}\label{GWmodel}

A field theory on Euclidean Moyal spacetime can be formulated in terms
of the Moyal $\star$-product, which is a nonlocal associative multiplication usually defined as
\begin{equation}\label{moyal}
    (\phi\star \psi)(x):=e^{i\,\partial^{^\phi}\Theta\partial^{^\psi}}\,\phi(x)\psi(x)\,.
\end{equation}
The scalar functions $\phi$ and $\psi$ depend on $x\in\mathbb{R}^d$;
$\partial^{^\phi}$ and $\partial^{^\psi}$ denote their gradients,
respectively\footnote{To avoid cluttering, throughout this article we omit writing
spacetime indices. Therefore, for any two vectors $u,v\in\mathbb{R}^d$ we write $uv$ to represent the usual scalar product $\sum_{a=1}^d u_a v_a$ and for a given matrix $M\in{\mathbb{R}^{d\times d}}$ we write $uMv:=\sum_{a,b=1}^d M_{ab}u_av_b$. Also, when it leads to no confusion, we omit the arguments of functions to ease the presentation.}. The real matrix $\Theta$ is skew-symmetric and characterizes the noncommutativity of spacetime. According to expression \eqref{moyal} the commutator between coordinates reads $[x_i,x_j]_\star=x_i\star x_j-x_j\star x_i=2i\,\Theta_{ij}$. The conditions under which the exponential in this expression is well-defined are studied in \cite{Estrada:1989da}.

In the following, up to section \ref{degen}, we consider a non-degenerate matrix $\Theta$; in consequence, the dimension $d$ of spacetime is even. Furthermore, the skew-symmetric matrix $\Theta$ can be written --upon an appropriate rotation-- in block-diagonal form as the direct sum of $d/2$ skew-symmetric $2\times 2$ matrices. This representation implies that noncommutative $\mathbb{R}^d$ can be decomposed into $d/2$ noncommutative planes, that commute among themselves, in each of which noncommutativity is represented by a real parameter. For simplicity, we assume that these $d/2$ parameters coincide. As a consequence, $\Theta$ satisfies $\Theta^2=-\theta^2\,\mathbf{1}_d$, where $\theta\in\mathbb{R}$ is a single parameter that describes noncommutativity in $\mathbb{R}^{d}$. An example of the implications of different values of the noncommutativity in the $d/2$ planes is discussed in section \ref{conclu}.

In terms of the Fourier transform $\mathcal{F}$,
\begin{equation}\label{foutra}
    \tilde{\phi}(p):=\mathcal{F}\{\phi\}(p)=\int_{\mathbb{R}^d} dx\,e^{-ipx}\phi(x)\,,
\end{equation}
the Moyal product reads
\begin{equation}\label{moyal-fourier}
    \mathcal{F}\{\phi\star \psi\}(p)=\int_{\mathbb{R}^d} d\tilde{p}'\,e^{-i p\Theta p'}\,\tilde{\phi}(p-p')\tilde{\psi}(p')\,,
\end{equation}
where\footnote{We will frequently use the following notation: a tilde on a function represents its Fourier transform $\mathcal{F}$, whereas a tilde on a differential denotes a factor $(2\pi)^{-d}$.} $d\tilde{p}':=dp'/(2\pi)^d$.

It is convenient to denote by $L(\phi)$ and $R(\phi)$ the left- and right-Moyal multiplication by $\phi$, that is
\begin{equation}
    L(\phi)\psi:=\phi\star \psi\qquad R(\phi)\psi:=\psi\star\phi\,,
\end{equation}
which, after a Bopp shift, can be written as
\begin{align}
    L(\phi)\psi(x)&=\phi(x+i\Theta \partial)\psi(x)\label{alt1}\,,\\
    R(\phi)\psi(x)&=\phi(x-i\Theta \partial)\psi(x)\label{alt2}\,.
\end{align}

Let us now consider the GW model for a real scalar field $\phi$ on $d$-dimensional Euclidean Moyal spacetime, which is defined by the Lagrangian\footnote{We use the parameter $\omega$ which is related by $\omega\,\theta=\Omega$ to the usual parameter $\Omega$ present in the original GW Lagrangian \cite{Grosse:2003nw}.}
\begin{equation}\label{lagrangian}
    \mathcal{L}=\frac{1}{2}(\partial\phi)^2+\frac{m^2}{2}\phi^2+
    \frac{\omega^2}{2}\,x^2\phi^2+\frac{\lambda}{4!}\phi^4_\star\,,
\end{equation}
where $\phi^4_\star:=\phi\star\phi\star\phi\star\phi$. The corresponding classical action is given by
\begin{equation}\label{action}
    S[\phi]=\int_{\mathbb{R}^d}dx\,\left(\frac{1}{2}\phi\, G\phi+\frac{\lambda}{4!}\phi^4_\star\right)\,,
\end{equation}
where
\begin{equation}\label{G}
  G:=-\partial^2+m^2+\omega^2x^2\,.
\end{equation}
The perturbative effective action, up to one-loop corrections, can be represented as
\begin{equation}\label{eff-action}
    \Gamma[\phi]=S[\phi]+\frac{1}{2}\log{\rm
      Det}\left\{\delta^2S\right\}\,.
\end{equation}
The (kernel of the) operator $\delta^2S$ is the second functional derivative of the action with respect to the field. From eq.\ \eqref{action} one obtains
\begin{align}\label{qf-operator}
  \delta^2S&=G+\frac{\lambda}{3!}\left[L(\phi^2_\star)+R(\phi^2_\star)+L(\phi)R(\phi)\right]\nonumber\\
           &=G+\frac{\lambda}{3!}\left[\phi^2_\star(x+i\Theta \partial)+
  \phi^2_\star(x-i\Theta \partial)+\phi(x+i\Theta \partial)\phi(x-i\Theta \partial)\right]\,.
\end{align}

A representation of the functional determinant in eq.\ \eqref{eff-action} is given by the Schwinger proper time approach, in which the one-loop corrections to the effective action are determined by the heat-trace of the operator $\delta^2S$
\begin{equation}\label{eff-action-pt}
    \Gamma[\phi]=S[\phi]-\frac{1}{2}\int_0^\infty \frac{d\beta}{\beta}\ {\rm Tr}\,
    e^{-\beta\left\{\delta^2S\right\}}\,.
\end{equation}
In this formulation the UV-divergencies of one-loop Feynman diagrams arise from the divergence of the integral in eq.\ \eqref{eff-action-pt} at $\beta\rightarrow 0$.

In the next section we compute the heat-trace in eq.\ \eqref{eff-action-pt} by using path integrals in the phase space $\mathbb{R}^{2d}$ of a point particle. However, we will establish results which are valid for more general nonlocal operators.

\section{The heat-trace of a nonlocal operator}\label{ht}

In this section, instead of considering the operator $\delta^2S$ given by eq.\ \eqref{qf-operator}, we study the heat-trace of the more general nonlocal operator
\begin{equation}\label{h}
    H:=-\partial^2+m^2+\omega^2x^2+V(x,-i\partial)\,.
\end{equation}
The operator $H$ acts on functions of $x\in\mathbb{R}^d$. The nonlocal operator $V$ is at this point arbitrary and will be later chosen to be the nonlocal part of $\delta^2S$.

The main idea of the WF is to interpret the heat-operator $e^{-\beta H}$ as the evolution operator, in Euclidean time $\beta$, of a quantum point particle moving on $\mathbb{R}^d$ subject to the (nonlocal) Hamiltonian $H$. Consequently, the heat-operator can be studied using the path integral methods of quantum mechanics. However, since the Hamiltonian given by eq.\ \eqref{h} is nonlocal, path integration must be implemented in phase space. In particular, since we are interested in the trace of the heat-operator we integrate on periodic trajectories in phase space. The usual discretization of these trajectories leads to the following path integral representation of the heat-trace
\begin{equation}\label{pi-h}
    {\rm Tr}\,e^{-\beta H}
        =e^{-\beta m^2}\mathcal{N}(\beta) \int\mathcal{D}q(t)\mathcal{D}p(t)\,e^{-\mathcal{S}[q(t),p(t)]}
        \ e^{-\beta\int_0^{1}dt\,V_W(\sqrt{\beta}q(t),p(t)/\sqrt{\beta})}
          \,,
\end{equation}
where
\begin{equation}\label{actionpoint}
  \mathcal{S}[q(t),p(t)]:=\int_0^1 dt\,\left\{p(t)^2-ip(t)\dot{q}(t)+\omega^2\beta^2q(t)^2\right\}
\end{equation}
is the Euclidean action of a harmonic oscillator in phase space $\mathbb{R}^{2d}$.

A few remarks are now in order. Firstly, note that we have made the following rescaling in the usual variables of the path integral: $t\rightarrow \beta t$, $q(t)\rightarrow \sqrt{\beta}q(t)$ and $p(t)\rightarrow p(t)/\sqrt{\beta}$ so that the trajectories $q(t),p(t)$ and the parameter $t$ are now dimensionless quantities. The normalization constant $\mathcal{N}(\beta)$ eventually accounts for this rescaling and will be determined later.

According to the midpoint prescription for path integrals, the function $V_W$ is obtained by replacing the operators $x$ and $-i\partial$ by $\sqrt{\beta}q(t)$ and $p(t)/\sqrt{\beta}$, respectively, in the Weyl ordered expression of the operator $V(x,-i\partial)$. This means that we must first symmetrize the nonlocal operator $V(x,-i\partial)$ with respect to $x$ and $-i\partial$ before inserting the trajectories $q(t),p(t)$. Nevertheless, an explicit calculation shows that in the case of the operator $\delta^2S$ given by eq.\ \eqref{qf-operator}, the corresponding nonlocal operator can be cast into its Weyl form without the introduction of any extra term \cite{Bonezzi:2012vr}.

Finally, the path integral in expression \eqref{pi-h} is performed on periodic trajectories $q(t),p(t)$, i.e., $q(0)=q(1)$ and $p(0)=p(1)$. As we mentioned, this boundary conditions are needed in order to obtain the trace of the heat-operator $e^{-\beta H}$. If one were interested in the heat-kernel $\langle x'|e^{-\beta H}|x\rangle$, then one should integrate on trajectories whose projection in configuration space satisfy the boundary conditions\footnote{An example of this kind of calculation can be found in Appendix \ref{app1}.} $q(0)=x$ and $q(1)=x'$. Of course, making $x=x'$ and integrating over $x$ would provide the heat-trace. However, for the calculation of the effective action the consideration of periodic boundary conditions will suffice.

As is usual, we denote the expectation value of a functional $f[q(t),p(t)]$ as
\begin{equation}\label{funct}
    \left\langle
      f[q(t),p(t)]\right\rangle:=\mathcal{Z}(\omega\beta)^{-1}\int\mathcal{D}q(t)\mathcal{D}p(t)\,e^{-\mathcal{S}[q(t),p(t)]} f[q(t),p(t)]\,,
\end{equation}
where we define
\begin{equation}\label{zob}
  \mathcal{Z}(\omega\beta):=\int\mathcal{D}q(t)\mathcal{D}p(t)\,e^{-\mathcal{S}[q(t),p(t)]}\,,
\end{equation}
so that $\langle 1\rangle=1$.

As before, trajectories $q(t),p(t)$ in expressions \eqref{funct} and \eqref{zob} satisfy periodic conditions at the boundaries $t=0$ and $t=1$. Using definition \eqref{funct} the heat-trace in eq.\ \eqref{pi-h} takes the simplified form
\begin{equation}\label{pi4}
        {\rm Tr}\,e^{-\beta H}
        =e^{-\beta m^2}\mathcal{N}(\beta)\mathcal{Z}(\omega\beta)\,
        \left\langle e^{-\beta\int_0^{1}dt\,V_W(\sqrt{\beta}q(t),p(t)/\sqrt{\beta})}\right\rangle
          \,.
\end{equation}
The normalization constant $\mathcal{N}(\beta)\mathcal{Z}(\omega\beta)$ can be readily determined from the well-known heat-trace of the $d$-dimensional harmonic oscillator. Indeed, if we put $m^2=V_W=0$ in eq.\ \eqref{pi4} we get
\begin{equation}
  {\rm Tr}\,e^{-\beta \{-\partial^2+\omega^2x^2\}}
        =\mathcal{N}(\beta)\mathcal{Z}(\omega\beta)=
        \left(\sum_{n=0}^\infty e^{-2\beta\omega(n+1/2)}\right)^d=
        \frac{1}{(2\sinh{\omega\beta})^{d}}
          \,.
\end{equation}
Next, we make an expansion of expression \eqref{pi4} in powers of the function $V_W$; this corresponds to an expansion in the number of vertices of the Feynman diagrams in a perturbative calculation of the one-loop effective action. If we introduce the Fourier transform $\tilde{V}_W(\cdot,\cdot)$ of $V_W(\cdot,\cdot)$ with respect to both of its arguments we finally get
\begin{align}{\label{pi3}}
        {\rm Tr}\,e^{-\beta H}
        &=\frac{e^{-\beta m^2}}{(2\sinh{\omega\beta})^{d}}
        \ \sum_{n=0}^\infty (-\beta)^n
        \int_{\mathbb{R}^{2d}}d\tilde{\sigma}_1d\tilde{\xi}_1\ldots
        \int_{\mathbb{R}^{2d}}d\tilde{\sigma}_nd\tilde{\xi}_n\,\times\nonumber\\[2mm]
        &\times\,\tilde{V}_W(\sigma_1,\xi_1)\ldots \tilde{V}_W(\sigma_n,\xi_n)
        \times K^{(n)}_\beta(\sigma_1,\ldots,\sigma_n;\xi_1,\ldots,\xi_n)
        \,,
\end{align}
where we have defined, for $n\geq 1$,
\begin{equation}\label{kn00}
    K^{(n)}_\beta(\sigma_1,\ldots,\sigma_n;\xi_1,\ldots,\xi_n)=
    \int_0^1dt_1\ldots\int_0^{t_{n-1}}dt_n\,
    \left\langle
    e^{i\int_0^1 dt\,\left\{k_n(t)p(t)+j_n(t)q(t)\right\}}
    \right\rangle
\end{equation}
and
\begin{align}
    k_n(t)&:=\beta^{-1/2}\,\sum_{i=1}^{n}\delta(t-t_i)\,\xi_i\,,\label{k}\\
    j_n(t)&:=\beta^{1/2}\,\sum_{i=1}^{n}\delta(t-t_i)\,\sigma_i\,.\label{j}
\end{align}
In expression \eqref{pi3} the term given by $n=0$ in the sum equals 1. The expectation value that determines $K^{(n)}_\beta$ in eq.\ \eqref{kn00} can be computed using the standard techniques for path integrals defined by a Gaussian measure. Indeed, the quadratic action of the point particle defined in eq.\ \eqref{actionpoint} can be written as
\begin{equation}\label{qufo}
  \mathcal{S}[q(t),p(t)]=\frac12\int_0^1dt\,
  \begin{pmatrix}
    p(t)&q(t)
  \end{pmatrix}
        D
  \begin{pmatrix}
    p(t)\\q(t)
  \end{pmatrix}\,,
\end{equation}
where the self-adjoint differential operator $D$ is given by
\begin{equation}
        D:=\begin{pmatrix}
                    2&-i\partial_t\\
                    i\partial_t&2\omega^2\beta^2
        \end{pmatrix}\,.
\end{equation}
The expectation value in expression \eqref{kn00} is now computed by the usual procedure of completing squares in the path integral. The result reads
\begin{equation}\label{z}
    \left\langle e^{i\int_0^1 dt \left\{k_n(t)p(t)+j_n(t)q(t)\right\}}\right\rangle=
    \exp{\left\{-\frac{1}{2}\int_0^1 dt\int_0^1 dt'\, \begin{pmatrix}
    k_n(t)&j_n(t)
    \end{pmatrix}
    G(t-t')
    \begin{pmatrix}
    k_n(t')\\j_n(t')
    \end{pmatrix}\right\}}\,,
\end{equation}
where $G(t-t')$ --the kernel of the inverse operator $D^{-1}$-- is given by
\begin{equation}\label{dkernel}
    G(\Delta)=\frac{1}{2\sinh{\omega\beta}}
    \begin{pmatrix}
            \omega\beta\,\cosh{[\omega\beta(2|\Delta|-1)]} &   i\,\epsilon(\Delta)\,\sinh{[\omega\beta(2|\Delta|-1)]}\\
            -i\,\epsilon(\Delta)\,\sinh{[\omega\beta(2|\Delta|-1)]} &   \frac{1}{\omega\beta}\,\,\cosh{[\omega\beta(2|\Delta|-1)]}
    \end{pmatrix}\,;
\end{equation}
the sign function $\epsilon(\cdot)$ is $\pm 1$ if its argument is positive or negative, respectively. The kernel $G(t-t')$ is symmetric and, as expected for periodic trajectories, depends only on the difference $\Delta:=t-t'$.

Replacing the sources $k_n(t),j_n(t)$, given by eqs.\ \eqref{k} and \eqref{j}, into eq.\ \eqref{z} we obtain
\begin{align}\label{kn}
    &K^{(n)}_\beta(\sigma_1,\ldots,\sigma_n;\xi_1,\ldots,\xi_n)=
    e^{-\frac{1}{4\omega\tanh{\omega\beta}}\, \stackrel[i]{}{\sum}(\omega^2\xi^2_i+\sigma^2_i)}
    \int_0^1dt_1\ldots\int_0^{t_{n-1}}dt_n\ \times\nonumber\\[2mm]
    &\times\,
    e^{-\frac{1}{2\omega\sinh{\omega\beta}}\, \stackrel[i<j]{}{\sum}
    \left\{\cosh{\left[\omega\beta(2|t_i-t_j|-1)\right]}\,(\omega^2\xi_i\xi_j+\sigma_i\sigma_j)
    +i\omega\sinh{\left[\omega\beta(2|t_i-t_j|-1)\right]}\,(\xi_i\sigma_j-\xi_j\sigma_i)\right\}}
    \,,
\end{align}
where the indices $i,j$ run from $1$ to $n$. This result, replaced in expression \eqref{pi3}, gives the heat-trace of the nonlocal operator given by eq.\ \eqref{h}.

\section{Effective action of the Grosse-Wulkenhaar model}\label{eaqw}

In the previous section we computed the heat-trace of the nonlocal operator given by eq.\ \eqref{h} in terms of an expansion in powers of $V$. In this section we will use this expansion to compute the effective action --given by eq.\ \eqref{eff-action-pt}-- of a real scalar field on Moyal spacetime with a quartic self-interaction and a harmonic oscillator background. Comparing expressions \eqref{qf-operator} and \eqref{h} we obtain for the corresponding function $V_W$ in phase space
\begin{equation}\label{qf-operator2}
    V_W(x,p)=\frac{\lambda}{3!}\left[\phi^2_\star(x-\Theta p)+
    \phi^2_\star(x+\Theta p)+
    \phi(x-\Theta p)\phi(x+\Theta p)\right]\,.
\end{equation}
As already mentioned, expression \eqref{qf-operator2} is justified by the fact that any operator that is a function of $x+i\Theta\partial$ and $x-i\Theta\partial$ can be written in its Weyl-ordered form without the introduction of extra terms \cite{Bonezzi:2012vr}.

Inserting expansion \eqref{pi3} into eq.\ \eqref{eff-action-pt} we obtain the following expression for the one-loop effective action corresponding to the real scalar field described by Lagrangian \eqref{lagrangian}:
\begin{align}\label{eff-action-pt2}
    \Gamma[\phi]&=S[\phi]+\frac{1}{2}\sum_{n=1}^\infty
        \int \prod_{i=1}^n \left\{d\tilde{\sigma}_id\tilde{\xi}_i\ \tilde{V}_W(\sigma_i,\xi_i)\right\}
        \int_{\Lambda^{-2}}^{\infty} d\beta\,
        \frac{e^{-\beta m^2}\,(-\beta)^{n-1}}{(2\sinh{\omega\beta})^{d}}
        \ K^{(n)}_\beta
        \,.
\end{align}
The functions $K^{(n)}_\beta$ can be read from expression \eqref{kn}. According to eq.\ \eqref{qf-operator2} the Fourier transform $\tilde{V}_W(\cdot,\cdot)$ is given by
\begin{align}\label{qf-operator2-ft}
    \tilde{V}_W(\sigma,\xi)&
    =\frac{\lambda}{3!}\,(2\pi)^d\left[\delta(\xi-\Theta\sigma)\ \mathcal{F}\left\{\phi^2_\star\right\}(\sigma)+
    \delta(\xi+\Theta\sigma)\ \mathcal{F}\left\{\phi^2_\star\right\}(\sigma)+\right.\nonumber\\
    &\left.\mbox{}+
    {\rm det}^{-1}(4\pi\Theta)\ \mathcal{F}\{\phi\}(\sigma/2-\Theta^{-1} \xi/2)\ \mathcal{F}\{\phi\}(\sigma/2+\Theta^{-1} \xi/2)\right]\,.
\end{align}
The symbol $\mathcal{F}$ in expression \eqref{qf-operator2-ft} represents the Fourier transform as a function on $\mathbb{R}^d$.

In expression \eqref{eff-action-pt2} we have introduced an UV-cutoff $\Lambda$ and we have removed the term corresponding to $n=0$ since it is a field-independent divergent quantity which represents the infinite volume contribution. Note also that IR-divergencies in the effective action, which would arise from integration at $\beta\rightarrow\infty$, are absent even in the massless case due to the exponentially decreasing factor $(\sinh{\omega\beta})^{-d}$ .

A term indexed by $n$ in eq.\ \eqref{eff-action-pt2} gives the contribution of the one-loop $2n$-point Green function to the effective action. In the next two sections we consider the terms corresponding to $n=1$ and $n=2$ in order to study one-loop corrections to the propagator and to the four-point function, respectively.

\section{Two-point function}\label{prop}

The contributions to the effective action which are quadratic in the field are given by the term corresponding to $n=1$ in eq.\ \eqref{eff-action-pt2}
\begin{equation}\label{2p-gf}
    \Gamma^{(2)}[\phi]=
        \frac{1}{2}
        \int_{\mathbb{R}^{2d}}d\tilde\sigma d\tilde\xi
        \ \tilde{V}_W(\sigma,\xi)
        \int_{\Lambda^{-2}}^{\infty} d\beta\,
        \frac{e^{-\beta m^2}}{(2\sinh{\omega\beta})^d}
        \ K^{(1)}_\beta(\sigma;\xi)
    \,,
\end{equation}
where
\begin{equation}
    K^{(1)}_\beta(\sigma;\xi)=e^{-\frac{1}{4\omega\tanh{\omega\beta}}\,(\omega^2\xi^2+\sigma^2)}\,.
\end{equation}

It is now convenient to consider each three terms in the Fourier transform $\tilde{V}_W$, given by eq.\ \eqref{qf-operator2-ft}, separately. The first two of them --which come from terms in $V_W$ that do not mix left- and right-Moyal multiplication-- correspond to the contributions of planar Feynman diagrams. On the other hand, the third term in eq.\ \eqref{qf-operator2-ft} --which comes from the term in $V_W$ that mixes left- and right-Moyal multiplication-- corresponds to the contribution of non-planar diagrams.

\subsection{Planar contributions}\label{planar}

The contributions of the first two terms in \eqref{qf-operator2-ft} to expression \eqref{2p-gf} coincide. This is clear from the fact that their only difference is the sign in the delta functions $\delta(\xi\pm\Theta\sigma)$ and that $K^{(1)}_\beta(\sigma;\xi)$ depends on $\xi^2$.

If we replace the first two terms of eq.\ \eqref{qf-operator2-ft} into eq.\ \eqref{2p-gf}, the planar corrections $\Gamma^{(2)}_{{\rm P}}[\phi]$ to the quadratic effective action read
\begin{align}\label{2p-gf-p2}
        \Gamma^{(2)}_{{\rm P}}[\phi]
        &=\frac{\lambda}{6}
        \int_{\mathbb{R}^{d}}d\tilde\sigma
        \ \mathcal{F}\left\{\phi^2_\star\right\}(\sigma)
        \int_{\Lambda^{-2}}^{\infty} d\beta\,
        \frac{e^{-\beta m^2}}{(2\sinh{\omega\beta})^d}
        \ e^{-\frac{1}{4\omega\tanh{\omega\beta}}\,(1+\omega^2\theta^2)\,\sigma^2}
        \nonumber\\[2mm]
        &=\frac{1}{2}\int_{\mathbb{R}^{d}}dx
        \ \Gamma^{(2)}_{{\rm P}}(x)\star\phi(x)\star\phi(x)\,,
\end{align}
where
\begin{align}
  \Gamma^{(2)}_{{\rm P}}(x)=\frac{\lambda}{3}\left\{\frac{\omega}{2\pi(1+\omega^2\theta^2)}\right\}^{d/2}
        \int_{\Lambda^{-2}}^{\infty} d\beta\,
        \frac{e^{-\beta m^2}}{(\sinh{2\omega\beta})^{d/2}}
        \ e^{-\frac{\omega\tanh{\omega\beta}}{1+\omega^2\theta^2}\,x^2}\,.\label{eff-back}
\end{align}
The function $\Gamma^{(2)}_{{\rm P}}(x)$ is UV-divergent due to the behavior of the integrand for small values of $\beta$. In the translational invariant case, $\omega=0$, $\Gamma^{(2)}_{{\rm P}}(x)$ is $x$-independent and its divergence is removed by the ordinary ($\theta$-independent) mass renormalization. However, for $\omega\neq 0$ the regularization of this divergence could also require frequency and field renormalization, according to the dimension of spacetime.

Let us consider, for instance, the planar contributions on a two-dimensional spacetime. For $d=2$, expression \eqref{2p-gf-p2} can be written as
\begin{equation}\label{2p-gf-c-2}
        \Gamma^{(2)}_{{\rm P}}[\phi]=\frac{1}{2}\int_{\mathbb{R}^{2}}dx\,
        \left\{ M_2^2\,\phi^2(x)+U_2(x)\star\phi(x)\star\phi(x)\right\}
        \,.
\end{equation}
As expected, expression \eqref{2p-gf-c-2} has only a logarithmic divergence \cite{Grosse:2003nw}, which is contained in the mass term
\begin{equation}
    M_2^2:=\frac{\lambda}{12\pi(1+\omega^2\theta^2)}
        \log{\left(\Lambda^{2}/\omega\right)}+{\rm (finite\ terms)}\,.
\end{equation}
The background $U_2(x)$ is UV-finite and can be represented as
\begin{equation}
    U_2(x):=\frac{\lambda}{12\pi(1+\omega^2\theta^2)}
        \int_{0}^{1}\frac{dt}{t}\,\left(\frac{1-t}{1+t}\right)^{m^2/2\omega}\,
        \left\{e^{-\frac{\omega}{1+\omega^2\theta^2}\,t\,x^2}-1\right\}\,.
\end{equation}
This background behaves as $\sim x^2$ for small $|x|$ so it contains a renormalization of the frequency $\omega$ --depending on the choice of renormalization prescriptions-- but which is nevertheless finite. In particular, for the massless case we obtain
\begin{equation}
    U_2(x)=-\frac{\lambda}{12\pi(1+\omega^2\theta^2)}
            \left\{
            \Gamma\left(0,\tfrac{\omega}{1+\omega^2\theta^2}\,x^2\right)+
            \log{\left(\tfrac{\omega}{1+\omega^2\theta^2}\,x^2\right)}+
            \gamma\right\}\qquad ({\rm for\ }m=0)\,,
\end{equation}
where $\Gamma(0,\cdot)$ is the incomplete gamma function and $\gamma$ is Euler-Mascheroni constant.

On the other hand, in higher dimensions the regularization of the quadratic effective action requires mass, field and frequency renormalization. Indeed, for $d=4$ expression \eqref{eff-back} takes the form
\begin{equation}
  \Gamma^{(2)}_{{\rm P}}(x)=\frac{\lambda}{48\pi^2}\frac{\omega}{(1+\omega^2\theta^2)^2}
        \int_{\tanh{\frac\omega{\Lambda^2}}}^{1}\ \frac{dt}{t^2}
        \ \frac{(1-t)^{\frac{m^2}{2\omega}+1}}{(1+t)^{\frac{m^2}{2\omega}-1}}
        \ \left\{1-\tfrac{\omega}{1+\omega^2\theta^2}\,t\,x^2\right\}+O(x^4)\,,\label{f4}
\end{equation}
which diverges quadratically as $\Lambda\rightarrow\infty$. The $O(x^4)$ term is UV-finite and positive. The first term between curly braces in expression \eqref{f4} is $x$-independent and is thus removed by an UV-divergent mass counterterm proportional to $\Lambda^2$, whereas the second term --proportional to $x^2$-- is logarithmically divergent and, when replaced in the second line of expression \eqref{2p-gf-p2}, leads to field and frequency renormalization\footnote{Note that $\int dx\,\{x^2\star\phi\star\phi\}=\int dx\,\{x^2\,\phi^2+\theta^2(\partial\phi)^2\}$.}. After this replacement we get for the field renormalization
\begin{equation}\label{z-1}
    Z=1+\frac{\lambda}{48\pi^2}\,\frac{\omega^2\theta^2}{(1+\omega^2\theta^2)^3}
    \left[\log{\left(\Lambda^2/\omega\right)}+{\rm (finite\ terms)}\right]\,,
\end{equation}
where $\phi_{\rm ren}(x)=Z^{-1/2}\,\phi(x)$. In consequence, as opposed to the commutative case\footnote{Note that, as expected, one-loop field renormalization vanishes for $\theta=0$ \cite{Goursac:2013kfa}.}, the field renormalization contains one-loop contributions. On the other hand, the same term in eq.\ \eqref{f4} gives for the renormalized frequency
\begin{equation}\label{w}
    \omega^2_{\rm ren}=\omega^2\left\{1-\frac{\lambda}{48\pi^2}\,\frac{1-\omega^2\theta^2}{(1+\omega^2\theta^2)^3}
    \left[\log{\left(\Lambda^2/\omega\right)}+{\rm (finite\ terms)}\right]\right\}\,.
\end{equation}
Expression \eqref{w} indicates that for $\omega\theta=1$ there is no (one-loop) frequency renormalization. This is a consequence of the existence of a fixed point of the renormalization group, due to its invariance under the Langman-Szabo duality \cite{Langmann:2002cc}. Expressions \eqref{z-1} and \eqref{w} coincide with the one-loop $\beta$- and $\gamma$-functions computed in \cite{Grosse:2004by}.

\subsection{Non-planar contribution}\label{non-planar}

The contribution of the third term in eq.\ \eqref{qf-operator2-ft} to the one-loop quadratic effective action reads
\begin{equation}\label{2p-gf-np}
        \Gamma^{(2)}_{{\rm NP}}[\phi]=\frac12
        \int_{\mathbb{R}^{2d}}d\tilde{p}d\tilde{p}'
        \,\tilde{\phi}(p)\tilde{\phi}(p')
        \ \Gamma^{(2)}_{{\rm NP}}(p,p')
        \,,
\end{equation}
with
\begin{align}
  \Gamma^{(2)}_{{\rm NP}}(p,p')
            &=\frac{\lambda}{6}
        \int_{0}^{\infty} \frac{d\beta}{({2}\sinh{\omega\beta})^d}\,
        e^{-\beta m^2-\frac{(p+p')^2+\omega^2\theta^2(p-p')^2}{4\omega\tanh{\omega\beta}}}
        \nonumber\\[2mm]
            &=\frac{\lambda}{12\omega}\,
        \Gamma(\tfrac{d}2+\tfrac{m^2}{2\omega})
        \ e^{-\frac{1}{4\omega}\,\left\{(p+p')^2+\omega^2\theta^2(p-p')^2\right\}}
        \ U\left(\tfrac{d}2+\tfrac{m^2}{2\omega},d;\tfrac{(p+p')^2+\omega^2\theta^2(p-p')^2}{2\omega}\right)
        \,,\label{2p-gamma}
\end{align}
where $U$ is Kummer's confluent hypergeometric function. The $\omega\rightarrow 0$ limit case of expression \eqref{2p-gamma} gives
\begin{equation}\label{2p-gf-np-ti}
        \Gamma^{(2)}_{\rm NP}(p,p')
        \rightarrow
        \frac{\lambda}{3}\,\pi^{\frac{d}2}m^{\frac{d}2-1}
        \ \frac{K_{d/2-1}(2m\theta |p|)}{(\theta |p|)^{\frac{d}2-1}}\ \delta(p+p')
\end{equation}
--where $K$ is the modified Bessel function-- which shows that, for $\omega=0$, the non-planar correction to the propagator diverges for small incoming momentum $p$ (UV/IR mixing problem \cite{Minwalla:1999px}).

The non-planar contribution given by eq.\ \eqref{2p-gf-np} represents a finite nonlocal correction to the background which can be written as
\begin{equation}
        \Gamma^{(2)}_{{\rm NP}}[\phi]
        =\frac12\int_{\mathbb{R}^{2d}}dxdx'\, \phi(x)\phi(x')\  U_{\rm NP}(x,x')
        \,,
\end{equation}
where the one-loop nonlocal background induced by the self-interaction is given by
\begin{equation}\label{polpot}
  U_{\rm NP}(x,x')=
  \frac{\lambda}{6}\,\frac{1}{(4\pi\theta)^d}\int_0^\infty d\beta\,\frac{e^{-\beta m^2}}{(\cosh{\omega\beta})^d}
    \ e^{-\frac{\tanh{\omega\beta}}{4\omega\theta^2}\left[(x-x')^2+\omega^2\theta^2(x+x')^2\right]}
  \,.
\end{equation}

\begin{figure}[t]
\centering
\begin{minipage}{.45\textwidth}
\centering
\includegraphics[height=45mm]{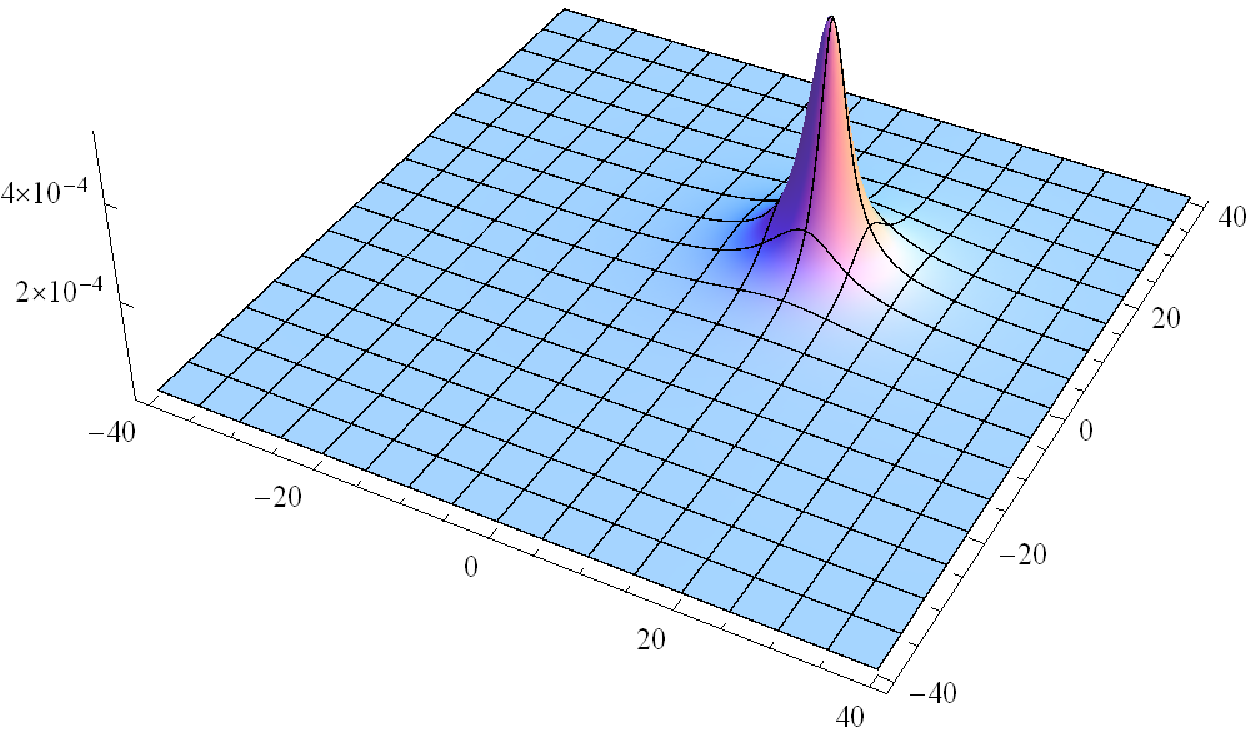}
\caption{Nonlocal background $U_{NP}(x,x')$ as a function of $x\in\mathbb{R}^2$ for $x'=(10,10)\in\mathbb{R}^2$, $\lambda=1$ and $\omega=0.1$ ($\sqrt{\theta}$ is taken as the unit length).}
\label{f2}
\end{minipage}
\hspace{0.05\textwidth}
\begin{minipage}{.45\textwidth}
\centering
\includegraphics[height=45mm]{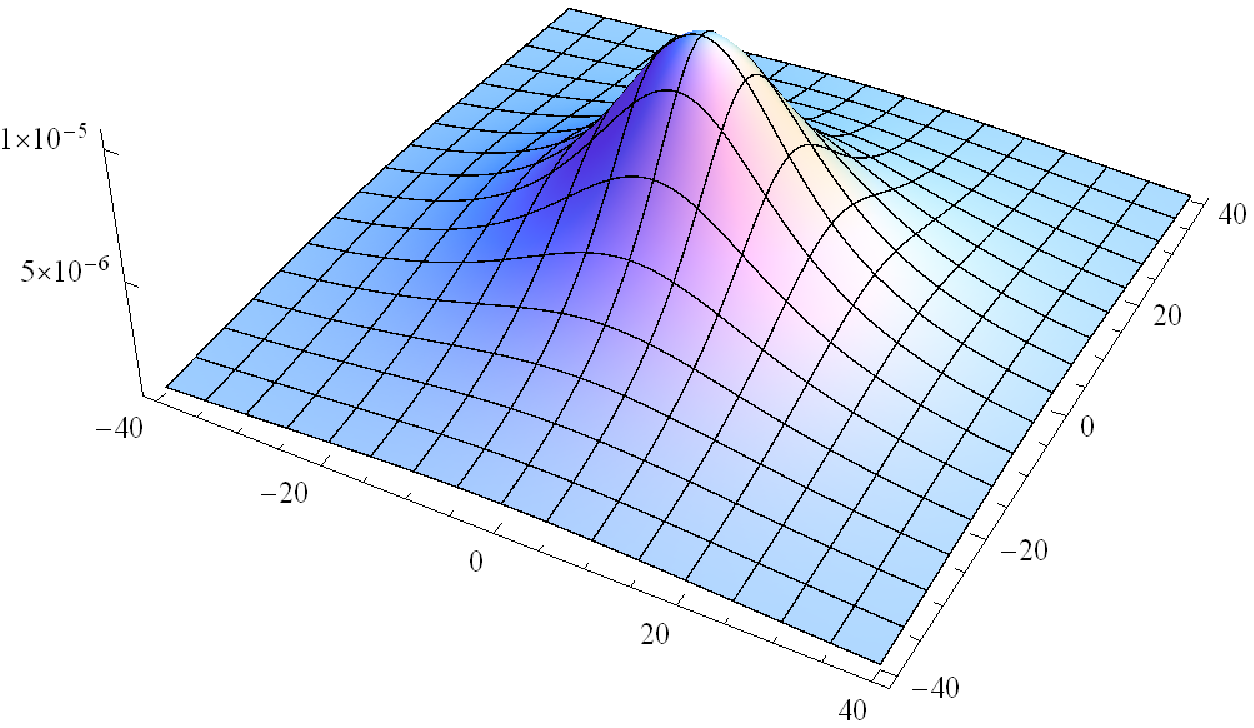}
\caption{Nonlocal background $U_{NP}(x,x')$ as a function of $x\in\mathbb{R}^2$ for $x'=(10,10)\in\mathbb{R}^2$, $\lambda=1$ and $\omega=1$ ($\sqrt{\theta}$ is taken as the unit length).}
\label{f3}
\end{minipage}
\end{figure}

For the particular case of a massless field ($m=0$) in two-dimensions ($d=2$) expression \eqref{polpot} reads
\begin{equation}
  U_{\rm NP}(x,x')=
  \frac{\lambda}{24\pi^2}\ \frac{\left(1-e^{-\frac{1}{4\omega\theta^2}\,\left[(x-x')^2+\omega^2\theta^2(x+x')^2\right]}\right)}%
  {\left[(x-x')^2+\omega^2\theta^2(x+x')^2\right]}
  \,.
\end{equation}
In figures \ref{f2} and \ref{f3} we display $U_{NP}(x,x')$ as a function of $x\in\mathbb{R}^2$ for two different values of $\omega\theta$. For small values of $\omega\theta$ the nonlocal background gets sharply peaked around $x=x'$, depending strongly on the distance $|x-x'|$ (fig.\ \ref{f2}). On the other hand, close to the self-dual point, where $\omega\theta=1$, the background depends separately on $|x|$ and $|x'|$, repelling the field away from the origin (fig.\ \ref{f3}). Finally, as $\omega\theta$ increases the function $U_{NP}$ gets concentrated around $x=-x'$.

\section{Four-point function}\label{fpf}

In this section we study the renormalization of the coupling constant $\lambda$. In order to do that, we consider the divergencies of the four-point function, which is given by the term corresponding to $n=2$ in formula \eqref{eff-action-pt2},
\begin{align}\label{eff-action-pt4}
    \Gamma^{(4)}[\phi]&=-\frac{1}{2}
        \int d\tilde{\sigma}_1d\tilde{\sigma}_2d\tilde{\xi}_1d\tilde{\xi}_2
        \ \tilde{V}_W(\sigma_1,\xi_1)
        \,\tilde{V}_W(\sigma_2,\xi_2)\,
        \int_{\Lambda^{-2}}^{\infty} d\beta\,\beta\,
        \frac{e^{-\beta m^2}}{(2\sinh{\omega\beta})^{d}}
        \ K^{(2)}_\beta
        \,.
\end{align}
The divergence of this expression as $\Lambda\rightarrow\infty$ comes from the leading term of $K^{(2)}_\beta$ in a small-$\beta$ expansion. This expansion is readily obtained from eq.\ \eqref{kn}
\begin{equation}\label{k2b0}
    K^{(2)}_\beta(\sigma_1,\sigma_2;\xi_1,\xi_2)\sim
    \,e^{-\frac{1}{4\omega^2\beta}\,\left\{\omega^2(\xi_1+\xi_2)^2+(\sigma_1+\sigma_2)^2\right\}}
    \int_{-1/2}^{1/2}dt\,\left(\tfrac12-t\right)\,e^{-it\,\left(\xi_1\sigma_2-\xi_2\sigma_1\right)}
    \,.
\end{equation}
This expression shows that upon integration $\xi_1,\sigma_1\sim\sqrt{\beta}$, so we can replace expansion \eqref{k2b0} into \eqref{eff-action-pt4} neglecting, to leading order in $\beta$, the contribution of the imaginary exponential. Since the remaining factors depend only on the sums $\xi_1+\xi_2$ and $\sigma_1+\sigma_2$, the integrals in \eqref{eff-action-pt4} give the convolution of the Fourier transforms $\tilde{V}_W$. As a consequence, in terms of $V_W(x,p)$ the result reads
\begin{align}\label{eff-action-4p}
    \Gamma^{(4)}[\phi]&=-\frac{1}{4}\frac{1}{(2\pi)^d}
        \int_{\mathbb{R}^{2d}} dxdp\ V^2_W(x,p)
        \int_{\Lambda^{-2}}^{\infty}d\beta\,\beta\,e^{-\beta m^2}\,\left(\frac{\omega\beta}{\sinh{\omega\beta}}\right)^d
        \,e^{-\beta\omega^2x^2-\beta p^2}
        +\ldots
\end{align}
where the dots represent sub-leading terms as $\beta\rightarrow 0$. Note that as long as the integral in phase space of $V^2_W(x,p)$ is finite, the exponentials can be removed and the integral in $\beta$ is finite as $\Lambda\rightarrow\infty$. This is the case for non-planar contributions, which are given by terms in $V^2_W(x,p)$ (see eq.\ \eqref{qf-operator2}) that depend on both independent combinations $(x+\Theta p)$ and $(x-\Theta p)$. On the contrary, the term $(\phi^2_\star(x-\Theta p))^2$ in $V^2_W(x,p)$ which depends only on $(x-\Theta p)$ would give a divergent phase space integral if the exponentials in \eqref{eff-action-4p} were removed. In fact, the exponential $\exp{(-\beta\omega^2x^2-\beta p^2)}$ guarantees a finite integral along the direction $(x+\Theta p)$; this integral provides an extra $\beta^{-d/2}$ term which, for $d\geq 4$, makes this contribution UV-divergent. The same argument holds for the term $(\phi^2_\star(x+\Theta p))^2$ in $V^2_W(x,p)$. In consequence, the divergent contributions to the four-point function in $d$ dimensions can be read from
\begin{equation}\label{4p-div}
    \Gamma^{(4)}[\phi]=-\frac{\lambda^2}{72}\frac{1}{(4\pi)^{d/2}}
        \int_{\mathbb{R}^d} dx
        \,\left\{\phi^2_\star(x)\right\}^2
        \int_{\Lambda^{-2}}^{\infty}d\beta\,\beta^{1-\tfrac d2}\left(\frac{\omega\beta}{\sinh{\omega\beta}}\right)^d
        \frac{e^{-\beta\left(m^2+\frac{\omega^2x^2}{1+\omega^2\theta^2}\right)}}{(1+\omega^2\theta^2)^{d/2}}
        +\ldots
\end{equation}
This expression is finite for $d=2$ so it only involves a finite renormalization of the coupling constant $\lambda$. On the contrary, for $d\geq 4$, one-loop contributions to the four-point function are divergent.

In the four-dimensional case the divergent contribution is $x$-independent and can therefore be removed by a coupling constant renormalization. Indeed, for $d=4$ expression \eqref{4p-div} can be written as
\begin{equation}\label{4p-div-d=4}
    \Gamma^{(4)}[\phi]=-\frac{\lambda^2}{1152\,\pi^2}
    \,\frac{1}{(1+\omega^2\theta^2)^2}
        \int_{\mathbb{R}^4} dx
        \ \phi^4_\star(x)
        \int_{\Lambda^{-2}}^{\omega^{-1}}\frac{d\beta}{\beta}\ +{\rm (finite\ terms)}\,,
\end{equation}
from where the renormalized coupling constant results
\begin{equation}\label{lambdar}
    \lambda_{\rm ren}=Z^2\,\lambda\left\{1-\frac{\lambda}{48\pi^2}\,\frac{1}{(1+\omega^2\theta^2)^2}
    \left[\log{\left(\Lambda^2/\omega\right)}+{\rm (finite\ terms)}\right]\right\}\,.
\end{equation}
The corresponding $\beta$-function (see also eq.\ \eqref{z-1}) is thus given by
\begin{equation}\label{betalambda}
    \beta_\lambda:=\Lambda\,\partial_\Lambda\lambda=\frac{\lambda^2_{\rm ren}}{24\pi^2}
    \ \frac{\left(1-\theta^2\omega^2_{\rm ren}\right)}{\left(1+\theta^2\omega^2_{\rm ren}\right)^3}\,,
\end{equation}
which, as expected, vanishes at the self-dual point \cite{Grosse:2004by}.

\section{Degenerate case}\label{degen}

To conclude, we show in this section how the WF works in the case of degenerate noncommutativity. As an example, we consider four-dimensional Euclidean spacetime parametrized by coordinates $(\bar{x},x)\in\mathbb{R}^4$, where $\bar{x}\in\mathbb{R}^2$ are commuting coordinates and $x\in\mathbb{R}^2$ describes a two-dimensional Moyal space where noncommutativity is characterized by
\begin{equation}\label{deg-theta}
    \Theta=\left(\begin{array}{cccc}
      0&&&\theta\\
      -\theta&&&0
    \end{array}
        \right)\,,
\end{equation}
with $\theta\in\mathbb{R}$. Let us study a real scalar field $\phi(\bar{x},x)$ in this deformed spacetime with Lagrangian
\begin{equation}\label{lagrangiandeg}
    \mathcal{L}=\frac{1}{2}(\bar{\partial}\phi)^2+\frac{1}{2}(\partial\phi)^2+\frac{m^2}{2}\phi^2+
    \frac{\omega^2}{2}\,x^2\phi^2+\frac{\lambda}{4!}\phi^4_\star\,,
\end{equation}
where $\bar{\partial}$ and $\partial$ are the two-dimensional gradients with respect to $\bar{x}$ and $x$, respectively. The $\star$-product in the quartic term is defined as in expression \eqref{moyal}, for $d=2$ and $\Theta$ given by eq.\ \eqref{deg-theta}, and therefore does not involve derivatives $\bar{\partial}$ with respect to the commutative coordinates $\bar{x}$. For this reason, the Lagrangian includes a harmonic oscillator term which only depends on the noncommutative coordinates $x$.

In order to determine the one-loop effective action, we proceed as in section \ref{ht} and evaluate the heat-trace of the nonlocal operator
\begin{equation}\label{hdeg}
    H:=-\bar{\partial}^2-\partial^2+m^2+\omega^2x^2+V(\bar{x},x,-i\partial)\,.
\end{equation}
Afterwards, the dependence of the Weyl symbol $V_W(\bar{x},x,p)$ of the operator $V(\bar{x},x,-i\partial)$ with respect to the coordinates $x,p\in \mathbb{R}^2$ will take the form given by eq.\ \eqref{qf-operator2}.

The usual path integral representation for the heat-trace reads
\begin{equation}\label{pi-h-deg}
    {\rm Tr}\,e^{-\beta H}
        =e^{-\beta m^2}\mathcal{N}(\beta) \int_{\mathbb{R}^2} d\bar{x}
        \int\mathcal{D}\bar{q}\mathcal{D}q\mathcal{D}p\ e^{-\mathcal{S}[\bar{q},q,p]}
        \ e^{-\beta\int_0^{1}dt\,V_W}
          \,,
\end{equation}
where $V_W:=V_W(\bar{x}+\sqrt{\beta}\,\bar{q}(t),\sqrt{\beta}\,q(t),p(t)/\sqrt{\beta})$. The point particle action can be separated as $\mathcal{S}[\bar{q},q,p]=\bar{\mathcal{S}}[\bar{q}]+\mathcal{S}[q,p]$ with
\begin{align}
  \bar{\mathcal{S}}[\bar{q}]&:=\frac14\int_0^1 dt\ \dot{\bar{q}}^2(t)\,,\label{actionpoint-bar}\\
  \mathcal{S}[q,p]&:=\int_0^1 dt\,\left\{p(t)^2-ip(t)\dot{q}(t)+\omega^2\beta^2q(t)^2\right\}\,.\label{actionpoint-deg}
\end{align}
Note that since $V_W$ does not depend on $\bar{\partial}$, conjugate momenta $\bar{p}(t)$ have been integrated out and the path integration in the commuting variables is performed only on  trajectories $\bar{q}(t)$ in configuration space. Trajectories $q(t),p(t)$ satisfy periodic boundary conditions whereas $\bar{q}(t)$ satisfies Dirichlet boundary conditions $\bar{q}(0)=\bar{q}(1)=0$; the full trace is obtained\footnote{The use of periodic trajectories $\bar{q}(t)$ does not simplify the computation of the trace due to the existence of a zero mode of the action \eqref{actionpoint-bar} under these boundary conditions.} after integration in the variable $\bar{x}$.

As in section \ref{ht}, the trace can be written as an expectation value defined as a path integral with measure given by the Gaussian actions \eqref{actionpoint-bar} and \eqref{actionpoint-deg},
\begin{equation}\label{pi4-deg}
        {\rm Tr}\,e^{-\beta H}
        =\frac{e^{-\beta m^2}}{16\pi\beta\sinh^2{\omega\beta}}\ \int_{\mathbb{R}^2} d\bar{x}
        \ \left\langle e^{-\beta\int_0^{1}dt\,V_W(\bar{x}+\sqrt{\beta}\,\bar{q}(t),\sqrt{\beta}\,q(t),p(t)/\sqrt{\beta})}\right\rangle
          \,.
\end{equation}
The normalization has been determined from the free case $V_W=0$. Next, we make an expansion in powers of the Fourier transform $\tilde{V}_W$,
\begin{align}{\label{pi3-deg}}
        {\rm Tr}\,e^{-\beta H}
        &=\frac\pi4\,\frac{e^{-\beta m^2}}{\beta\sinh^2{\omega\beta}}
        \ \sum_{n=1}^\infty (-\beta)^n
        \int_{\mathbb{R}^{6}}d\tilde{\bar{\sigma}}_1d\tilde{\sigma}_1d\tilde{\xi}_1\ldots
        \int_{\mathbb{R}^{6}}d\tilde{\bar{\sigma}}_n\tilde{\sigma}_nd\tilde{\xi}_n\,\times\nonumber\\[2mm]
        &\times\,\delta\left(\bar{\sigma}_1+\ldots+\bar{\sigma}_n\right)\,\tilde{V}_W(\bar{\sigma}_1,\sigma_1,\xi_1)\ldots \tilde{V}_W(\bar{\sigma}_n,\sigma_n,\xi_n)
        \times K^{(n)}_\beta
        \,,
\end{align}
where
\begin{equation}\label{kn00-deg}
    K^{(n)}_\beta=
    \int_0^1dt_1\ldots\int_0^{t_{n-1}}dt_n\,
    \left\langle
    e^{i\int_0^1 dt\,\left\{k_n(t)p(t)+j_n(t)q(t)\right\}}
    \right\rangle_{{\rm periodic}}\,
    \left\langle
    e^{i\int_0^1 dt\,\bar{j}_n(t)\bar{q}(t)}
    \right\rangle_{{\rm Dirichlet}}
\end{equation}
and
\begin{equation}\label{kdeg}
    k_n(t):=\beta^{-\frac12}\,\sum_{i=1}^{n}\delta(t-t_i)\,\xi_i\,,\quad
    j_n(t):=\beta^{\frac12}\,\sum_{i=1}^{n}\delta(t-t_i)\,\sigma_i\,,\quad
    \bar{j}_n(t):=\beta^{\frac12}\,\sum_{i=1}^{n}\delta(t-t_i)\,\bar{\sigma}_i\,.\quad
\end{equation}
In expression \eqref{pi3-deg} we have removed the field independent term corresponding to $n=0$. The delta-function in the sum $\sum_i \bar{\sigma}_i$ is a result of the integration in $\bar{x}$. The expectation value that determines $K^{(n)}_\beta$ in eq.\ \eqref{kn00-deg} has been separated into a path integration in phase space $q(t),p(t)$, with periodic boundary conditions, and a path integration over $\bar{q}(t)$, with Dirichlet boundary conditions. The former has already been computed in section \ref{ht} in terms of the Green functions for periodic boundary conditions. The result reads
\begin{align}\label{kn-deg}
    &\left\langle e^{i\int_0^1 dt\,\left\{k_n(t)p(t)+j_n(t)q(t)\right\}}\right\rangle_{{\rm periodic}}=
    e^{-\frac{1}{4\omega\tanh{\omega\beta}}\, \stackrel[i]{}{\sum}(\omega^2\xi^2_i+\sigma^2_i)}\ \times\nonumber\\[2mm]
    &\times\,
    e^{-\frac{1}{2\omega\sinh{\omega\beta}}\, \stackrel[i<j]{}{\sum}
    \left\{\cosh{\left[\omega\beta(2|t_i-t_j|-1)\right]}\,(\omega^2\xi_i\xi_j+\sigma_i\sigma_j)
    +i\omega\sinh{\left[\omega\beta(2|t_i-t_j|-1)\right]}\,(\xi_i\sigma_j-\xi_j\sigma_i)\right\}}
    \,.
\end{align}
The expectation value with respect to trajectories in the commutative sector of spacetime is in turn given by
\begin{equation}\label{kn-bar}
    \left\langle e^{i\int_0^1 dt\,\bar{j}_n(t)\bar{q}(t)}\right\rangle_{{\rm Dirichlet}}=
    e^{-\beta\, \stackrel[i,j]{}{\sum}\,g(t_i,t_j)\bar{\sigma}_i\bar{\sigma}_j}\,,
\end{equation}
where $g(t,t')$ is the Green function under Dirichlet boundary conditions\footnote{That is, the kernel of the inverse of $-\partial^2_t$ with Dirichlet conditions at $t=0$ and $t=1$.}
\begin{equation}
  g(t,t')=\frac12\left\{
    t(1-t')+t'(1-t)-|t-t'|
  \right\}\,.
\end{equation}
The indices in the sums of eqs.\ \eqref{kn-deg} and \eqref{kn-bar} take the values $i,j=1,\ldots,n$.

The contributions to the one-loop effective action which are quadratic in the field are then given by the term corresponding to $n=1$ in eq.\ \eqref{pi3-deg}
\begin{equation}\label{2p-gf-deg}
    \Gamma^{(2)}[\phi]=
        \frac{1}{32\pi}
        \int_{\mathbb{R}^{4}}d\tilde\sigma d\tilde\xi\ \tilde{V}_W(0,\sigma,\xi)
        \int_{\Lambda^{-2}}^{\infty} d\beta\,
        \frac{e^{-\beta m^2}}{\beta\sinh^2{\omega\beta}}
        \ K^{(1)}_\beta
    \,,
\end{equation}
where
\begin{equation}
    K^{(1)}_\beta=e^{-\frac{1}{4\omega\tanh{\omega\beta}}\,(\omega^2\xi^2+\sigma^2)}\,.
\end{equation}

The planar contribution to the two-point function can be written as
\begin{align}\label{2p-gf-p2-deg}
        \Gamma^{(2)}_{{\rm P}}[\phi]
        =\frac{1}{2}\int_{\mathbb{R}^{4}}d\bar{x}dx\ \Gamma^{(2)}_{{\rm P}}(x)\star\phi(\bar{x},x)\star\phi(\bar{x},x)\,,
\end{align}
where
\begin{equation}
  \Gamma^{(2)}_{{\rm P}}(x)=\frac{\lambda}{24\pi^2}\,\frac{\omega}{(1+\omega^2\theta^2)}
        \int_{\Lambda^{-2}}^{\infty} \frac{d\beta}{\beta}\, \frac{e^{-\beta m^2}}{\sinh{2\omega\beta}}
        \ e^{-\frac{\omega\tanh{\omega\beta}}{1+\omega^2\theta^2}\,x^2}\,.\label{eff-action-deg}
\end{equation}
The result is similar to the one obtained for the planar contributions in the four-dimensional non-degenerate case (see eqs.\ \eqref{eff-back} and \eqref{f4}): $\Gamma^{(2)}_{{\rm P}}(x)$ has an $x$-independent UV-divergence proportional to $\Lambda^2$, which is removed by mass renormalization, and a logarithmic divergence, which is removed by the following field and frequency renormalization:
\begin{align}
    Z&=1+\frac{\lambda}{48\pi^2}\,\frac{\omega^2\theta^2}{(1+\omega^2\theta^2)^2}
    \left[\log{\left(\Lambda^2/\omega\right)}+{\rm (finite\ terms)}\right]\,,\label{zdeg}\\
    \omega^2_{\rm ren}&=\omega^2\left\{1-\frac{\lambda}{48\pi^2}\,\frac{1-\omega^2\theta^2}{(1+\omega^2\theta^2)^2}
    \left[\log{\left(\Lambda^2/\omega\right)}+{\rm (finite\ terms)}\right]\right\}\,.\label{wdeg}
\end{align}
Since two of the four coordinates commute, these expressions show an extra $(1+\omega^2\theta^2)$ factor compared to eqs.\ \eqref{z-1} and \eqref{w}. Note also that one-loop corrections to the kinetic term only introduce divergencies in the noncommutative directions, i.e.\ $\frac12(\partial\phi)^2$, and not in the derivatives with respect to the commutative variables $\frac12(\bar\partial\phi)^2$. As a consequence, one-loop divergencies are removed not only by the field renormalization $Z$ but after the introduction of a new parameter which takes this asymmetry between commutative and noncommutative directions into account \cite{Grosse:2012my}.

Finally, non-planar contributions, given by the third term in expression \eqref{qf-operator2-ft}, read
\begin{equation}\label{2p-gf-np2-deg}
        \Gamma^{(2)}_{{\rm NP}}[\phi]
        =\frac{1}{2}\int_{\mathbb{R}^{6}}d\bar{x}dxdx'\,\phi(\bar{x},x)\,\phi(\bar{x},x')\ U_{{\rm NP}}(x,x')\,,
\end{equation}
where
\begin{equation}
  U_{{\rm NP}}(x,x')=\frac{\lambda}{384\pi^3}\,\frac{1}{\theta^2}
        \int_{\Lambda^{-2}}^{\infty} \frac{d\beta}{\beta}\, \frac{e^{-\beta m^2}}{\cosh^2{\omega\beta}}
        \ e^{-\frac{\tanh{\omega\beta}}{4\omega\theta^2}\,\left[(x-x')^2+\omega^2\theta^2(x+x')^2\right]}\,.\label{eff-back-deg}
\end{equation}
This result is to be compared with the background obtained in the non-degenerate case (see eq.\ \eqref{polpot} for $d=4$). As has already been shown, non-planar contributions in the non-degenerate case are finite, regardless of the dimension of spacetime; however, in the degenerate case the effective background $U_{{\rm NP}}(x,x')$ given by eq.\ \eqref{eff-back-deg} has a logarithmic UV-divergence
\begin{equation}\label{lognonpladeg}
  U_{{\rm NP}}(x,x')=\frac{\lambda}{384\pi^3}\,\frac{1}{\theta^2}\,\log{\Lambda^2/\omega}+{\rm (finite\ terms)}\,.
\end{equation}
This type of divergence cannot be removed by any of the parameters in the Lagrangian \eqref{lagrangiandeg}. As a consequence, one-loop renormalization in the degenerate case requires introducing the new nonlocal interaction term
\begin{equation}
        \frac{1}{2}\,\frac{\kappa^2}{\theta^2}\int_{\mathbb{R}^{6}}d\bar{x}dxdx'\,\phi(\bar{x},x)\,\phi(\bar{x},x')\,,
\end{equation}
as described in \cite{Grosse:2008df}. The logarithmic divergence given by eq.\ \eqref{lognonpladeg}, which is $x$-independent, is therefore removed by a renormalization of the parameter $\kappa$. The $\beta$-functions derived from eqs.\ \eqref{zdeg}, \eqref{wdeg} and \eqref{lognonpladeg} are in complete agreement with the results of \cite{Grosse:2012my}.

\section{Conclusions}\label{conclu}

We have obtained a master formula --eq.\ \eqref{eff-action-pt2}-- for the one-loop effective action of the GW model in $d$-dimensional Euclidean spacetime from which planar and non-planar contributions to the $n$-point Green function can be readily computed. Note that the same expression \eqref{eff-action-pt2} can also be applied to self-interactions of the form $\lambda\,\phi_\star^N$ by choosing the appropriate $\tilde{V}_W(\sigma,\xi)$.

We studied $2$-point and $4$-point Green functions of the GW model, where non-planar contributions are identified by the presence of both left- and right-Moyal multiplications or, accordingly, by simultaneous factors depending on $\xi\pm\Theta \sigma$ in the products of $\tilde{V}_W$. From the planar contributions the appropriate $\beta$-functions were obtained. Finally, also the case of degenerate noncommutativity was considered, where the need to introduce a new counterterm is made explicit and the existence of a fixed point of the renormalization group is verified.

For the derivation of this master formula we needed to compute the heat-trace of the nonlocal operator which gives the quantum fluctuations of the scalar field. Nevertheless, we considered a more general class of nonlocal operators --eq.\ \eqref{h}-- for which the use of worldline techniques gives a representation of the heat-kernel in terms of path integrals in the phase space of a point particle.

As it has been shown, the WF is particularly convenient to study nonlocal operators, whose symbols are non-polynomial functions of momenta. Indeed, in path integration in phase space, momenta are replaced by ordinary trajectories so non-standard dependence on momenta appear, in this sense, in an equal footing with the dependence on the coordinates. Therefore, we considered operators which are written as a quadratic expression in momenta and coordinates plus an interaction term which is formally an arbitrary function of both of them. For this class of nonlocal operators we computed the asymptotic expansion of the heat-trace --eq.\ \eqref{pi3}-- in powers of the proper time or, equivalently, in powers of this nonlocal interaction term.

It is worth pointing out a remark regarding our use of periodic trajectories in the implementation of the WF. It has been shown that, in ordinary commutative flat space, path integrations in the worldline approach give the same result if performed using either periodic or Dirichlet boundary conditions \cite{Fliegner:1997rk}; in fact, the effective Lagrangians differ in total derivative terms but after integration in infinite spacetime one obtains an unambiguous result for the effective action. In general, computations under periodic boundary conditions are simplified by translation invariance; however, either under periodic or Dirichlet conditions an integration in spacetime --in the ``center of mass'' zero mode or in the common intersection of the loops, respectively-- has to be performed. On the other hand, for field theories on curved spacetime the use of periodic trajectories might be misleading since coordinate covariance is lost in the computation of local quantities \cite{Schalm:1998ix}. In the case considered in this article one can check by explicit calculation that the same expression \eqref{pi3} for the heat-trace would have been obtained --even though through a longer route-- if we integrated over trajectories satisfying Dirichlet boundary conditions. However, the use of periodic trajectories in the GW model is much more convenient not only because of translation invariance (in the worldline) but also due to the presence of the harmonic oscillator background which removes the zero mode (in target phase space) and thus avoids the spacetime integration in the heat-trace calculation.

As regards the application of this expansion to the GW model, let us mention that one could have considered, instead of the isotropic harmonic oscillator background $\omega^2\,x^2$ in the nonlocal operator \eqref{h}, a general quadratic form $M_{ij}\,x_ix_j$. After an appropriate rotation, this background can be written as the sum of one-dimensional oscillators with frequencies $\omega_i$, with $i=1,\ldots,d$. In this case, the same procedure of section \ref{ht} can be performed to compute the heat-trace, the result being the same expression \eqref{pi3} with $\omega$ understood as the diagonal matrix\footnote{Of course, expressions such as $(2\sinh{\omega\beta})^d$ should be replaced by ${\rm det}\{2\sinh{\omega\beta}\}$.} ${\rm diag}\,(\omega_1,\ldots,\omega_d)$. This trivial generalization is necessary for the application to the GW model in the case in which the skew-symmetric matrix $\Theta$ --in its $2\times 2$ block-decomposition-- is characterized by $d/2$ different parameters $\theta_n$, with $n=1,\ldots,d/2$ (see the discussion before eq.\ \eqref{foutra}). In this anisotropic model, the quadratic background in the Lagrangian \eqref{lagrangian} is replaced by $\tfrac12\sum_{i}\omega_i^2\,x_i^2$, where $\omega_{2n-1}=\omega_{2n}=\Omega/\theta_n$ and $\Omega$ is a dimensionless parameter \cite{Grosse:2003nw}. After this modification one can reproduce our calculations and find, as in eq.\ \eqref{w}, that the $\beta$-function corresponding to the interaction with this background vanishes at $\Omega=1$. The explicit calculation also shows the importance of the anisotropy of the background for removing UV-divergencies. In other words, this anisotropy in the quadratic background cancels the anisotropy in the noncommutativity parameters; otherwise, one-loop divergencies would not be cancelled, e.g., by the single parameter $Z$ in the field renormalization. Though seemingly trivial, this fact illustrates a remarkable property of the GW model: not any confining background would give a noncommutative renormalizable theory --free of UV/IR mixing-- as one could have naively expected\footnote{On the other hand, it is the anisotropic oscillator background that maps to the kinetic term under the Langmann-Szabo duality $p\leftrightarrow \Theta^{-1}\,x$.}.

\medskip

The WF has been successfully applied to a wide variety of field theories in ordinary commutative space \cite{Schubert:2001he,Bastianelli:2006rx}, in particular to compute effective actions of spin 1 fields \cite{Strassler:1992zr,Reuter:1996zm}, where the simplifications afforded by worldline methods are particularly significant --both in abelian and nonabelian gauge field theories (see e.g.\ \cite{Bastianelli:2005vk,Dai:2008bh,Pawlowski:2008xh,Ahmadiniaz:2012ie}). In the last years several NC gauge theories have been proposed but their renormalizability has not been proved yet \cite{Blaschke:2009rb} mainly because the usual techniques to prove renormalization either break gauge invariance or have to be adapted to nonlocal theories. A natural open question is whether the procedure described in this article can be implemented to study Green functions in these models. Research along this line is currently in progress. In point of fact, we consider the phase space representation could be useful for studying other nonlocal theories (not necessarily noncommutative) as well as higher-derivative field theories.

\section*{Acknowledgments}

The authors acknowledge support from CONICET, Argentina. Financial support from CONICET (PIP 0681), ANPCyT (PICT 0605) and UNLP (X615) is also acknowledged. PP gratefully acknowledges the organizers of the Workshop on Spinning Particles in Quantum Field Theory (San Crist\'obal de las Casas, M\'exico; 2013) and UCMEXUS-CONACYT CN-12-564 for supporting his participation in this meeting from which the present work has very much benefited. PP also acknowledges Prof.\ Christian Schubert for very useful comments regarding the implementation of different boundary conditions in the worldline formalism.

\appendix

\section{Mehler Kernel}\label{app1}

In order to illustrate how the WF can also be used for computing local heat-kernels, in this appendix we compute the free propagator
\begin{equation}
  \Delta(p,p'):=\int_{\mathbb{R}^{2d}}dxdx'\,e^{-ipx-ip'x'}\,\langle x'|G^{-1}|x\rangle\,;
\end{equation}
that is, the Fourier transform of the kernel of the inverse operator $G^{-1}$ (see eq.\ \eqref{G}), which satisfies
\begin{equation}\label{ed}
  (-\omega^2\partial^2+m^2+p^2)\Delta(p,p')=(2\pi)^d\,\delta(p+p')\,.
\end{equation}

First, we express the inverse operator $G^{-1}$ in terms of the heat-operator $e^{-\beta\, G}$, so that we can write
\begin{equation}
  \Delta(p,p')=\int_{\mathbb{R}^{2d}}dxdx'\,e^{-ipx-ip'x'}\,
  \int_{0}^\infty d\beta
  \,\langle x'|e^{-\beta G}|x\rangle
  \,.
\end{equation}
Next, we write a path integral representation of the heat-kernel
\begin{align}\label{a1}
  \Delta(p,p')&=\int_{\mathbb{R}^{2d}}dxdx'\,e^{-ipx-ip'x'}
  \ \times\nonumber\\
  &\times
  \int_{0}^\infty d\beta\,e^{-\beta m^2}\int\mathcal{D}q(t)\mathcal{D}p(t)
  \,e^{-\int_0^{\beta}
          dt\,\left\{p^2(t)-ip(t)\dot{q}(t)+\omega^2q^2(t)\right\}}
  \,.
\end{align}
In the path integral of expression \eqref{a1} trajectories $p(t)$ do not satisfy any prescribed boundary condition, whereas the trajectories $q(t)$ satisfy $q(0)=x$ and $q(\beta)=x'$. It is convenient to perform the path integral on perturbations around the classical solutions $q_0(t),p_0(t)$, which are the trajectories that minimize the action\footnote{Note that we do not perform the scalings in $p(t),q(t),t$ which were used in the text.}
\begin{equation}
    \mathcal{S}_0[q(t),p(t)]=\int_0^{\beta}dt\,\left\{p^2(t)-ip(t)\dot{q}(t)+\omega^2q^2(t)\right\}
\end{equation}
with the aforementioned boundary conditions. These trajectories are given by
\begin{align}
  q_0(t)&=\frac{1}{\sinh{2\omega\beta}}\left[x'\,\sinh{2\omega t}+x\,\sinh{2\omega (\beta-t)}\right]\,,\\
  p_0(t)&=\frac{i\omega}{\sinh{2\omega\beta}}\left[x'\,\cosh{2\omega t}-x\,\cosh{2\omega (\beta-t)}\right]\,.
\end{align}
If we make the shift $q(t)\rightarrow q_0(t)+q(t)$ and $p(t)\rightarrow p_0(t)+p(t)$ in expression \eqref{a1} we obtain
\begin{align}\label{a30}
  \Delta(p,p')&=\int_{0}^\infty d\beta\,e^{-\beta m^2}\int_{\mathbb{R}^{2d}}dxdx'\,e^{-ipx-ip'x'}
  \ \times\nonumber\\
  &\times
  \,e^{-\frac{\omega}{2\sinh{2\omega\beta}}\left\{(x^2+x'^2)\cosh{2\omega\beta}-2xx'\right\}}
  \,
  \int\mathcal{D}q(t)\mathcal{D}p(t)
  \,e^{-\int_0^{\beta}
  dt\,\left\{p^2(t)-ip(t)\dot{q}(t)+\omega^2q^2(t)\right\}}
  \,.
\end{align}
The first exponential in the second line of eq.\ \eqref{a30} is the exponential of the action at the classical trajectories in phase space $\mathcal{S}_0[q_0(t),p_0(t)]$ and the path integral in $q(t),p(t)$ represents the contribution of the quantum fluctuations. Note that the integral is now performed on trajectories whose projection in configuration space satisfy the homogeneous Dirichlet conditions $q(0)=q(\beta)=0$. This path integral can be interpreted as the transition amplitude at coinciding points $x=x'=0$ in Euclidean time $\beta$ of a $d$-dimensional harmonic oscillator. Therefore it can be computed from the explicit form of the eigenfunctions of the harmonic oscillator Hamiltonian in terms of Hermite polynomials $H_n(\cdot)$. The result reads
\begin{align}\label{n2}
  \int\mathcal{D}q(t)\mathcal{D}p(t)
  \,e^{-\int_0^{\beta}
  dt\,\left\{p^2(t)-ip(t)\dot{q}(t)+\omega^2q^2(t)\right\}}
  =\langle x'=0|\,e^{-\beta\left\{-\partial^2+\omega^2x^2\right\}}\,|x=0\rangle\nonumber\\
  =\left(\sum_{n=0}^\infty e^{-2\beta\omega(n+\frac12)}
  \frac{\sqrt{\omega}}{2^n n!\sqrt{\pi}}\,H^2_n(0)\right)^d=
  \left(\frac{\omega}{2\pi\sinh{2\omega\beta}}\right)^{d/2}\,.
\end{align}
Inserting eq.\ \eqref{n2} into eq.\ \eqref{a30} we finally obtain
\begin{equation}\label{a3}
  \Delta(p,p')
  =\left(\frac{2\pi}{\omega}\right)^{d/2}
  \int_{0}^\infty d\beta\,\frac{e^{-\beta m^2}}{(\sinh{2\omega\beta})^{d/2}}
  e^{-\frac{1}{2\omega\sinh{2\omega\beta}}\left\{
  (p^2+p'^2)\cosh{2\omega\beta}+2pp'
  \right\}}
  \,,
\end{equation}
that is the Mehler kernel in Fourier space \cite{mehler}.



\begin{thebibliography}{99}

\bibitem{Schubert:2001he}
  C.~Schubert,
  ``Perturbative quantum field theory in the string inspired formalism,''
  Phys.\ Rept.\  {\bf 355} (2001) 73.

\bibitem{Bastianelli:2006rx}
  F.~Bastianelli, P.~van Nieuwenhuizen,
  ``Path integrals and anomalies in curved space,''
  Cambridge, UK: Univ. Pr. (2006).

\bibitem{Douglas:2001ba}
  M.~R.~Douglas, N.~A.~Nekrasov,
  ``Noncommutative field theory,''
  Rev.\ Mod.\ Phys.\  {\bf 73} (2001) 977-1029.

\bibitem{Szabo:2001kg}
  R.~J.~Szabo,
  ``Quantum field theory on noncommutative spaces,''
  Phys.\ Rept.\  {\bf 378} (2003) 207-299.

\bibitem{Bonezzi:2012vr}
  R.~Bonezzi, O.~Corradini, S.~A.~Franchino Vi\~nas and P.~A.~G.~Pisani,
  ``Worldline approach to noncommutative field theory,''
  J.\ Phys.\ A {\bf 45} (2012) 405401.

\bibitem{Minwalla:1999px}
  S.~Minwalla, M.~Van Raamsdonk, N.~Seiberg,
  ``Noncommutative perturbative dynamics,''
  JHEP {\bf 0002 } (2000)  020.

\bibitem{Vassilevich:2003yz}
  D.~V.~Vassilevich,
  ``Noncommutative heat kernel,''
  Lett.\ Math.\ Phys.\  {\bf 67} (2004) 185-194.

\bibitem{Grosse:2003nw}
  H.~Grosse and R.~Wulkenhaar,
  ``Renormalization of phi**4 theory on noncommutative R**2 in the matrix base,''
  JHEP {\bf 0312} (2003) 019.

\bibitem{Grosse:2004yu}
  H.~Grosse and R.~Wulkenhaar,
  ``Renormalization of phi**4 theory on noncommutative R**4 in the matrix base,''
  Commun.\ Math.\ Phys.\  {\bf 256} (2005) 305.

\bibitem{Langmann:2002cc}
  E.~Langmann and R.~J.~Szabo,
  ``Duality in scalar field theory on noncommutative phase spaces,''
  Phys.\ Lett.\ B {\bf 533} (2002) 168.

\bibitem{Buric:2009ss}
  M.~Buric and M.~Wohlgenannt,
  ``Geometry of the Grosse-Wulkenhaar Model,''
  JHEP {\bf 1003} (2010) 053.

\bibitem{Rivasseau:2005bh}
  V.~Rivasseau, F.~Vignes-Tourneret and R.~Wulkenhaar,
  ``Renormalization of noncommutative phi**4-theory by multi-scale analysis,''
  Commun.\ Math.\ Phys.\  {\bf 262} (2006) 565.

\bibitem{Disertori:2006nq}
  M.~Disertori, R.~Gurau, J.~Magnen and V.~Rivasseau,
  ``Vanishing of Beta Function of Non Commutative Phi**4(4) Theory to all orders,''
  Phys.\ Lett.\ B {\bf 649} (2007) 95.

\bibitem{Rivasseau:2007qda}
  V.~Rivasseau,
  ``Non-commutative Renormalization,''
  Prog.\ Math.\ Phys.\  {\bf 53} (2007) 19.

\bibitem{Grosse:2012uv}
  H.~Grosse and R.~Wulkenhaar,
  ``Self-Dual Noncommutative $\phi^4$ -Theory in Four Dimensions is a Non-Perturbatively Solvable and Non-Trivial Quantum Field Theory,''
  Commun.\ Math.\ Phys.\  {\bf 329} (2014) 1069.

\bibitem{Grosse:2014lxa}
  H.~Grosse and R.~Wulkenhaar,
  ``Solvable 4D noncommutative QFT: phase transitions and quest for reflection positivity,''
  arXiv:1406.7755 [hep-th].

\bibitem{Grosse:2014nza}
  H.~Grosse and R.~Wulkenhaar,
  ``Construction of the $\Phi^4_4$-quantum field theory on noncommutative Moyal space,''
  arXiv:1402.1041 [math-ph].

\bibitem{Sfondrini:2010zm}
  A.~Sfondrini and T.~A.~Koslowski,
  ``Functional Renormalization of Noncommutative Scalar Field Theory,''
  Int.\ J.\ Mod.\ Phys.\ A {\bf 26} (2011) 4009.

\bibitem{Estrada:1989da}
  R.~Estrada, J.~M.~Gracia-Bondia, J.~C.~Varilly,
  ``On Asymptotic expansions of twisted products,''
  J.\ Math.\ Phys.\  {\bf 30} (1989) 2789.

\bibitem{Goursac:2013kfa}
  A.~de Goursac,
  ``Renormalization of the commutative scalar theory with harmonic term to all orders,''
  Annales Henri Poincare {\bf 14} (2013) 2025.

\bibitem{Grosse:2004by}
  H.~Grosse and R.~Wulkenhaar,
  ``The beta function in duality covariant noncommutative phi**4 theory,''
  Eur.\ Phys.\ J.\ C {\bf 35} (2004) 277.

\bibitem{Grosse:2012my}
  H.~Grosse and M.~Wohlgenannt,
  ``Degenerate noncommutativity,''
  Eur.\ Phys.\ J.\ C {\bf 72} (2012) 2153.

\bibitem{Grosse:2008df}
  H.~Grosse and F.~Vignes-Tourneret,
  ``Quantum field theory on the degenerate Moyal space,''
  J.\ Noncommut.\ Geom.\  {\bf 4} (2010) 555.

\bibitem{Fliegner:1997rk}
  D.~Fliegner, P.~Haberl, M.~G.~Schmidt and C.~Schubert,
  ``The Higher derivative expansion of the effective action by the string inspired method. Part 2,''
  Annals Phys.\  {\bf 264} (1998) 51.

\bibitem{Schalm:1998ix}
  K.~Schalm and P.~van Nieuwenhuizen,
  ``Trace anomalies and the string inspired definition of quantum - mechanical path integrals in curved space,''
  Phys.\ Lett.\ B {\bf 446} (1999) 247.

\bibitem{Strassler:1992zr}
  M.~J.~Strassler,
  ``Field theory without Feynman diagrams: One loop effective actions,''
  Nucl.\ Phys.\ B {\bf 385} (1992) 145.

\bibitem{Reuter:1996zm}
  M.~Reuter, M.~G.~Schmidt and C.~Schubert,
  ``Constant external fields in gauge theory and the spin 0, 1/2, 1 path integrals,''
  Annals Phys.\  {\bf 259} (1997) 313.

\bibitem{Bastianelli:2005vk}
  F.~Bastianelli, P.~Benincasa and S.~Giombi,
  ``Worldline approach to vector and antisymmetric tensor fields,''
  JHEP {\bf 0504} (2005) 010.

\bibitem{Dai:2008bh}
  P.~Dai, Y.~t.~Huang and W.~Siegel,
  ``Worldgraph Approach to Yang-Mills Amplitudes from N=2 Spinning Particle,''
  JHEP {\bf 0810} (2008) 027.

\bibitem{Pawlowski:2008xh}
  J.~M.~Pawlowski, M.~G.~Schmidt and J.~H.~Zhang,
  ``On the Yang-Mills two-loop effective action with wordline methods,''
  Phys.\ Lett.\ B {\bf 677} (2009) 100.

\bibitem{Ahmadiniaz:2012ie}
  N.~Ahmadiniaz, C.~Schubert and V.~M.~Villanueva,
  ``String-inspired representations of photon/gluon amplitudes,''
  JHEP {\bf 1301} (2013) 132.

\bibitem{Blaschke:2009rb}
  D.~N.~Blaschke, E.~Kronberger, A.~Rofner, M.~Schweda, R.~I.~P.~Sedmik and M.~Wohlgenannt,
  ``On the Problem of Renormalizability in Non-Commutative Gauge Field Models: A Critical Review,''
  Fortsch.\ Phys.\  {\bf 58} (2010) 364.

\bibitem{mehler}
    F.G.~Mehler,
     ``Ueber die Entwicklung einer Function von beliebig vielen Variablen nach Laplaceschen Functionen h\"oherer Ordnung,''
     J.\ Reine Angew.\ Math.\ {\bf 66} (1866) 161.


\end{thebibliography}
\end{document}